\documentclass[showpacs,twocolumn]{revtex4}
\usepackage{graphicx}
\usepackage{bm}
\usepackage{amsmath}
\usepackage{amsfonts}
\usepackage{amssymb}

\begin{document}

\title{Screening of nuclear pairing in nuclear and neutron matter}

\author{Caiwan Shen$^{1,2}$, U. Lombardo$^{3,4}$, P. Schuck$^{5}$}
\affiliation{
$^{1}$China Institute of Atomic Energy, P.O.Box 275(18), Beijing 102413, China \\
$^{2}$Center of Theoretical Nuclear Physics, National Laboratory of HIC, Lanzhou, China\\
$^{3}$LNS-INFN, Via Santa Sofia 44, I-95123 Catania, Italy  \\
$^{4}$Dipartimento di Fisica, Via Santa Sofia 64, I-95123 Catania, Italy \\
$^{5}$Institut de Physique Nucl\'{e}aire, Universit\'{e} Paris-Sud, F-91406
      Orsay Cedex, France}


\begin{abstract}
The screening potential in the $^1S_0$ and $^3S_1$ pairing channels
in neutron and nuclear matter in
different approximations is discussed. It is found that the vertex corrections to
the potential are much stronger in nuclear matter than in neutron matter.

\end{abstract}
\maketitle

\section{Introduction}

The question of the screening of nuclear pairing is a long standing problem.
In the past it has mostly been considered in neutron matter (see
Ref.~\cite{schul} for a list of references). However, quite recently also
conjectures about the possibility of anti-screening, i.e., enhancement of
pairing in finite nuclei due to surface vibrations have been put forward
\cite{bort}. In Ref. \cite{shen} we also contributed to the question of the
screening of the bare nucleon-nucleon pairing force in pure neutron matter and
found consistently with the results of other authors that this polarization
quenches quite strongly the neutron pairing in the $S=0$, $T=1$ channel.
However, as far as we know, similar investigations have never been carried out
in nuclear matter. Being interesting in its own right, our interest here
is to try to make a qualitative link to finite nuclei.

Despite the fact that the validity of the local density
approximation (LDA) may be questioned due to the large coherent
length of the Cooper pairs, it still yields, for example, for the
pair correlation energy, semi-quantitative agreement with
microscopic Hartree-Fock-Bogolyubov (HFB) calculations in finite
nuclei \cite{schuc}. Let us therefore resume our present knowledge
about nuclear pairing in different channels: as already mentioned,
practically all calculations in the spin singlet channel of
neutron matter indicate that polarization quenches pairing. What
about $S=0$, $T=1$ pairing in finite nuclei? In an important very
recent contribution to this subject it has been shown \cite{x1}
that the bare nucleon-nucleon (NN) force yields already $\sim50\%$
of the measured gap in the tin isotopes. It must be mentioned that
in that calculation the usual effective density dependent $k$-mass
with $m^{\ast }/m\sim0.7$ at saturation has been employed. This is
in line with the common lowest order approach to pairing in
nuclear physics where for example with the successful Gogny D1S
force \cite{gogny} in HFB calculations implicitly also the
$k$-mass is used \cite{schuc}. It further should be mentioned that
within the same approach and specifically with the Gogny D1S force
the $S=0$, $T=1$ gap in symmetric nuclear matter is very close to
the one obtained with the bare force for densities
$\rho\leq\rho_{0}/5$ whereas it drops off quite a bit slower for
densities $\rho>\rho_{0}/5$. At saturation ($\rho=\rho_{0}$), the bare
force yields almost negligible gap whereas D1S still yields
$\Delta\sim 0.5$ MeV (see Fig.~2 in Ref. \cite{x2}). In a LDA
picture it is therefore not unreasonable that with the bare force
the gap is about a factor 2 smaller than the one from this
D1S-force \cite{x1}. The D1S force has therefore pairing
properties which are not very far from the bare one. Remember that
the weak coupling formula predicts an exponential dependence of
the gap on the force (which however may be somewhat questioned in
finite nuclei due to the discreteness of the spectrum). Indeed it
has been shown recently that a renormalized bare force
($v_{\text{low-k}}$) which is phase shift equivalent for low
energies ($\leq300$ MeV) has shape and magnitude very similar to
the Gogny D1S force \cite{kuo}. The fact that D1S is close to the
bare force in the $S=0$, $T=1$ pairing channel has also been
noticed by Bertsch et al.\cite{x4} in their investigation of the
neutron halo in $^{11}$Li.

Concluding this discussion, we can expect that medium
renormalization of the bare $S=0$, $T=1$ pairing interaction
should yield some additional attraction in nuclear
matter. This is indeed what is claimed to be the case in
Ref.\cite{sorb} and it is also what we will find in present
study of infinite nuclear matter, employing exactly the same
approach as the one used before in neutron matter \cite{shen}.
However, as we will see later, the effect is dramatically strong.
In finite nuclei, for instance in the presently very
actively studied $N\simeq Z$ nuclei, we have the further very
interesting neutron-proton ($np$) $S=1$, $T=0$ pairing channel. It is
generally believed that the pairing force in this channel should
be of similar strength, may be a little stronger, than the one in
the $S=0$, $T=1$ channel \cite{lane}. In Refs.\cite{x2,bbl,bbs} the
gap in the $S=1$, $T=0$ channel, calculated with the bare force (and
$k$-mass) in infinite symmetric nuclear matter, is given as a
function of density, see e.g., Fig.~1 of Ref.\cite{bbs}
and Fig.~1 of Ref.\cite{x2}. Because of the
stronger attraction of the bare force in the deuteron channel the
$np$ gap turns out to be much larger than the
neutron-neutron ($nn$) or proton-proton ($pp$) one. For example at maximum
the $np$ gap is about 8 MeV (!)
whereas it is 2.5 MeV for the $nn$ gap. Even
at saturation $\Delta_{np}$ still is of the order of 2 MeV (!).
Clearly, the $np$ gap in finite nuclei would turn out much too
large, if the bare force was employed. Screening therefore should
give additional repulsion in the $S=1$, $T=0$. We therefore have
from the experimental side, and keeping in mind that LDA should at
least give the right trend when going from the infinite
homogeneous case to finite nuclei, definite predictions what the
inclusion of polarization in a nuclear matter calculation should
give, as a trend, in the $S=0$, $T=1$ and $S=1$,$T=0$ channels: additional
attraction (over the bare interaction) in the former and repulsion
in the latter. We will see in how far these expectations are
fulfilled by the calculations.

\section{Bubble screening}

The superfluid phase of a homogenous system of fermions is characterized by
the pairing field $\Delta_{k}(\omega)$, which is the solution of the
generalized gap equation\cite{nozi,mig,risch}
\begin{equation}
\Delta_{\bm{k}}(\omega)=\sum\limits_{k^{\prime}}\int\frac{d\omega^{\prime}%
}{2\pi i}\mathcal{V}_{\bm{k},\bm{k}^{\prime}}(\omega,\omega^{\prime}%
)F_{\bm{k}^{\prime}}(\omega^{\prime}),
\end{equation}
where $\mathcal{V}$ is the sum of all irreducible  NN interaction
terms and $F_{k}(\omega)$ is the anomalous propagator. Dealing with a
strongly correlated Fermi system one expects the medium
corrections to play a crucial role. The gap equation by itself
embodies the full class of particle-particle (p-p) ladder diagrams
just taking the bare interaction for $\mathcal{V}$ \cite{mig}. The
same kind of correlations are incorporated in the propagators if
the self-energy is approximated by the Brueckner-Hartree-Fock
(BHF) mean field \cite{cugn}. From this side additional p-p
corrections are unlikely to improve the predictions since the
pairing rapidly vanishes at high density. On the contrary,
particle-hole (p-h) correlations should
play an important role at low density where
the pairing gap is expected to take the largest value. But again p-h
contributions have to be treated on equal footing, in the vertex
corrections as well as in the self-energy in view of possible
strong cancellations. This was the main concern of the study
presented in Ref.\cite{shen} for $^{1}S_{0}$ $nn$ pairing in neutron
matter, further extended in a preliminary
work to nuclear matter in \cite{sorb}.
The expansion of the interaction block $\mathcal{V}$
and the self-energy $\Sigma$ were both truncated to the second order
in the interaction. This approximation turns out to be reasonable
for the self-energy, whereas it is only indicative for the trend of
the screening effects due to the interaction. In fact it was
shown that the screening on the $^{1}S_{0}$ $nn$ pairing is
quite different according to whether the medium is made out of
neutrons or neutron plus protons. Quantitative predictions require
that the full RPA bubble series, at least, should be summed up in
order to calculate the screening interaction, being aware that
even this approximation could be not enough to remove the
singularity associated to the low-density instability of nuclear
matter.

\begin{figure}[htbp]
\begin{center}
\includegraphics[width=85mm]{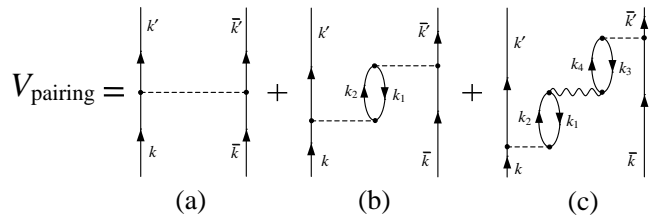}
\end{center}
\caption{Pairing interaction with screening in the RPA
approximation. The dashed lines represent the Gogny interaction,
the wiggly line the p-h residual interaction resummed to all orders. All
vertices are to be understood as anti-symmetrised (not shown explicitly).}
\label{diagram}
\end{figure}

According to the preceding discussion, we extend the study of
Ref.\cite{shen} including in $\mathcal{V}$ the full bubble series.
It is convenient to develop the latter as shown in
Fig.~\ref{diagram}, where the block (c) is the correction to the
results of Ref.\cite{shen}. The splitting of the bubble series
into the diagrams (b) and (c) enable us to disjoin the
consideration of the two external vertices connecting one particle
line with one hole line from the full p-h vertex (wiggly line). So
doing, the former two vertices can be calculated taking exactly
into account their full momentum dependence, whereas the latter
can be restricted to transitions around the Fermi momentum. In
fact it is well known that the magnitude of the gap is quite
sensitive to high momenta transitions, i.e. the short range part
of the nuclear force \cite{cugn}, whereas p-h excitations are only
important near the Fermi surface.

The p-h bubble series is summed up by the Bethe-Salpeter equation
\cite{nozi}\begin{eqnarray}
{V}_{\mathcal{S},\mathcal{T}}^{\rm RPA}(\bm{q}_{1},\bm{q}_{2},q)
&\!\!=\!\!&{V}_{\mathcal{S},\mathcal{T}}(\bm{q}_{1},\bm{q}_{2},q)+\int
\frac{d^{4}q_{3}}{(2\pi)^{4}}\Lambda(q_{3},q)\nonumber\\
&&  \times{V}_{\mathcal{S},\mathcal{T}}(\bm{q}_{1},\bm{q}_{3},q%
){V}_{\mathcal{S},\mathcal{T}}^{\rm RPA}(\bm{q}_{3},\bm{q}_{2},q).
\end{eqnarray}
Here $q=(\omega,\bm{q})$ and $\mathcal{S},\mathcal{T}$ denote the
total spin and isospin in the p-h channel. $\Lambda
(q_{3},q)=-iG(q_{3})G(q+q_{3})$ is the polarization
insertion\cite{fet}. Since, as we said, only small energy-momentum
transfers significantly contribute to the pairing interaction
around the Fermi energy, it is likely a valid approximation to
replace the p-h residual interaction with the effective interaction
$V$ expressed in terms of Landau parameters:
\begin{equation}
v=f+g\,(\bm{\sigma}\cdot\bm{\sigma}^{\prime})+f^{\prime}\,(\bm{\tau}%
\cdot\bm{\tau}^{\prime})+g^{\prime}\,(\bm{\sigma}\cdot\bm{\sigma}^{\prime
})\,(\bm{\tau}\cdot\bm{\tau}^{\prime}).
\end{equation}
The remarkable advantage of this approximation is that,
it makes it easy to sum up the bubble expansion of the p-h residual
interaction (for a review see Ref.\cite{back}). An additional
advantage is that one is incorporating the short-range p-p
correlations in the p-h effective interaction. In terms of the
Landau parameters one gets
\begin{eqnarray}
N(0){V}_{\mathcal{S},\mathcal{T}}^{\rm RPA}(q) &  =
&\frac{F}{1+\Lambda(q)F}+\frac
{G}{1+\Lambda(q)G}(\bm{\sigma}\cdot\bm{\sigma}^{\prime})\nonumber\\
&&
+\frac{F^{\prime}}{1+\Lambda(q)F^{\prime}}(\bm{\tau}\cdot\bm{\tau
}^{\prime})\nonumber\\
&&
+\frac{G^{\prime}}{1+\Lambda(q)G^{\prime}}(\bm{\sigma}\cdot\bm{\sigma
}^{\prime})\,(\bm{\tau}\cdot\bm{\tau}^{\prime}),
\end{eqnarray}
where $F=N(0)f$, $G=N(0)g$, $F'=N(0)f'$, $G'=N(0)g'$
and $N(0)$ is the level density on the Fermi
surface. $\Lambda(q)$ is the Lindhard function resulting from the
integration of the polarization insertion \cite{fet}(see
Appendices). This interaction will be adopted as vertex insertion
(wiggly line) in the two-bubble diagram as shown in Fig.~\ref{diagram}c.
So it represents the missing RPA correction in the one-bubble
approximation for the screening, already calculated with the Gogny
force \cite{shen}. Of course, this choice entails that the Landau
parameters must be calculated using the same interaction, i.e. the
Gogny force in our calculations, adopted for the two external
interaction vertices. The Landau parameters with the Gogny force
D1~\cite{footnote}, after RPA bubble summation at a
given density, depend on energy and momentum transfer
via the Lindhard functions and their asymptotic values coincide
with the previous values at the same density.

At this point let us again discuss our choice of the Gogny force
in the vertices (dashed lines) of Fig.~\ref{diagram}. In principle
a good approximation to the dashed lines in Fig.~1b and 1c would
be a microscopic $G$-matrix. This is well known from many body theory.
We here replace the $G$-matrix by the phenomenological Gogny force
which has precisely been adjusted to a $G$-matrix calculation
(see Ref.\cite{footnote}). More questionable is our use of the Gogny
force in the Born term of Fig.~1a. In principle in Fig.~1a the bare
force should be taken and the use of the Gogny force is for pure
convenience here because a bare force scatters to very high energies.
However, for better quantitative comparison of the Born terms,
Fig.~1a, and the screening terms, Fig.~1b and 1c, rather the
use of an effective low energy force like $v_{\rm low-k}$ \cite{kuo},
which is phase shift equivalent with the bare force at low
energies, would be appropriate. Since for us such a renormalised
$v_{\rm low-k}$ is not available in this work and because we know
from \cite{kuo} that $v_{\rm low-k}$ and Gogny forces are quite similar,
we feel entitled to use the Gogny force for the Born term of Fig.~1a.
In view of the fact that, as will be shown below, the contributions
of the screening terms are not at all small compared with the Born
term, we feel that a slight inaccuracy in the evaluation of the Born term
will not at all have strong consequences for the conclusions which
will be drawn in this paper. In this respect we also should mention that
in the $^1S_0$ channels, the density dependent part of the Gogny force
drops out in the Born term. However, this is not the case in the
$^3S_1$ channel.

Different screening mechanisms come into play according to whether
the medium is nuclear matter or neutron matter, and also whether
one has pairing between like or unlike particles. Denoting by $S$
and $T$ total spin and total isospin in the p-p channel, the
diagrams of Fig.~\ref{diagram} are written in the spin and isospin
representation
\begin{equation}
\mathcal{V}_{ST}=\mathcal{V}_{ST}^{(a)}+\mathcal{V}_{ST}^{(b)}+\mathcal{V}_{ST}^{(c)},
\label{rpaeqn}
\end{equation}
where
$$
\mathcal{V}_{ST}^{(a)}  =
{\displaystyle\sum_{\sigma_{i}\tau_{i}}} \left\langle
k\bar{k}|V_{ST}|k^{\prime}\bar{k}^{\prime}\right\rangle C_{ST},
$$
\begin{widetext}
\begin{eqnarray*}
\mathcal{V}_{ST}^{(b)} &  =&
{\displaystyle\sum_{\sigma_{i}\tau_{i}S_{i}T_{i}k_{i}}}
(-1)^{\sigma_{1}+\tau_{1}}C(\sigma-\sigma_{1};\sigma^{\prime}\sigma_{2}
|S_{1})C(\tau-\tau_{1};\tau^{\prime}\tau_{2}|T_{1})
\nonumber  \times
C(\sigma_{2}\bar{\sigma};-\sigma_{1}\bar{\sigma}^{\prime}
|S_{2})C(\tau_{2}\bar{\tau};-\tau_{1}\bar{\tau}^{\prime}|T_{2})
\nonumber\\
& & \times\left\langle
kk_{1}|V_{S_{1}T_{1}}|k^{\prime}k_{2}\right\rangle \left\langle
k_{2}\bar{k}|V_{S_{2}T_{2}}|k_{1}\bar{k}^{\prime}\right\rangle
\Lambda(k_{1},k_{2})C_{ST},
\nonumber\\
\mathcal{V}_{ST}^{(c)} &  =&
{\displaystyle\sum_{\sigma_{i}\tau_{i}S_{i}T_{i}k_{i}}}
(-1)^{\sigma_{1}+\tau_{1}+\sigma_{3}+\tau_{3}}C(\sigma-\sigma_{1}
;\sigma^{\prime}\sigma_{2}|S_{1})C(\tau-\tau_{1};\tau^{\prime}\tau_{2}
|T_{1}) \times
C(\sigma_{4}\bar{\sigma};-\sigma_{3}\bar{\sigma}^{\prime}
|S_{3})C(\tau_{4}\bar{\tau};-\tau_{3}\bar{\tau}^{\prime}|T_{3})
\nonumber\\
&&  \times\left\langle
kk_{1}|V_{S_{1}T_{1}}|k^{\prime}k_{2}\right\rangle
V_{\mathcal{S}_{2}\mathcal{T}_{2}}^{\rm RPA}\left\langle k_{4}\bar{k}
|V_{S_{3}T_{3}}|k_{3}\bar{k}^{\prime}\right\rangle
\Lambda(k_{1},k_{2} )\Lambda(k_{3},k_{4})C_{ST},\nonumber
\end{eqnarray*}
\end{widetext}
in which
$$
C_{ST}=C(\sigma\bar{\sigma};\sigma^{\prime}\bar{\sigma}^{\prime}|S)C(\tau
\bar{\tau};\tau^{\prime}\bar{\tau}^{\prime}|T)
$$
and
\begin{eqnarray*}
C(\sigma_{1}\sigma_{2};\sigma_{3}\sigma_{4}|S) &\!\!=\!\!& \left\langle
{{\tfrac{1}{2}} }\sigma_{1}{{\tfrac{1}{2}}}
\sigma_{2}\right\vert  \left.  S\ \sigma
_{1}+\sigma_{2}\right\rangle
 \\ && \times
\left\langle {{\tfrac{1}{2}}}\sigma_{3} {{\tfrac{1}{2}}}\sigma_{4}\right\vert
\left. S\ \sigma_{3}+\sigma _{4}\right\rangle
\delta_{\sigma_{1}+\sigma_{2},\sigma_{3}+\sigma_{4}}\:.
\end{eqnarray*}
 It is worth noticing that the $V^{\rm RPA}$ insertion is depending only
on the total spin and isospin in the p-h channel, whereas the two
external interactions are expressed in the spin and isospin
coupling mixed representation of p-p and p-h channels.

\begin{figure*}[thb]
\begin{center}
\includegraphics[width=14cm]{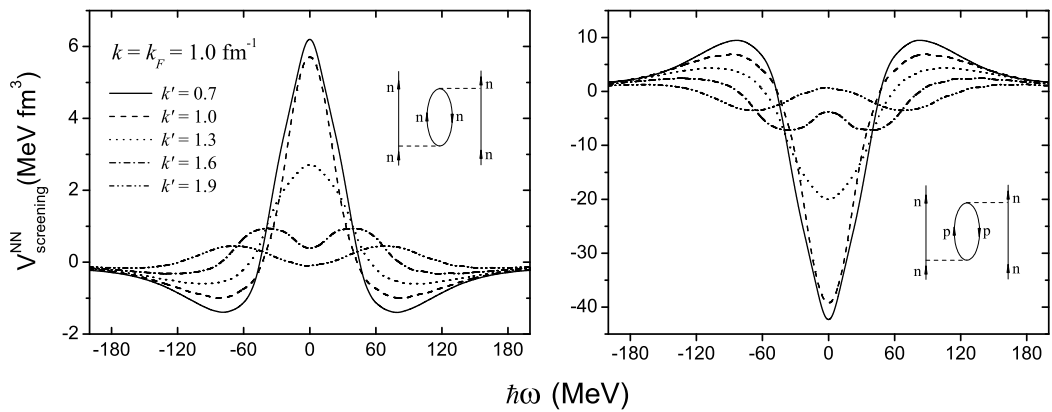}
\end{center}
\caption{Splitting of the $^1S_0$ pairing screening potential
${\mathcal V}_{k_Fk'}(\omega)$ in nuclear matter. Left (right) panel is
for the neutron (proton) p-h one-bubble insertion. The Fermi momentum
is fixed to 1.0 fm$^{-1}$.}
\label{v2-1s0-np}%
\end{figure*}

Nuclear matter within the RPA treatment of p-h residual interaction
suffers from a mechanical instability signaled by the well known
singularity of the compression modulus at $F_{0}=-1$. It occurs in
a density domain where pairing gap is large, and its influence on
the screening compromises any quantitative prediction. In neutron
matter the instability would not appear, but for several
interactions including Gogny D1, $F_{0}$ approaches dangerously
$-1$ so making doubtful any estimate. This drawback has been cured
by the Babu-Brown theory \cite{babu,back} where the class of
bubble diagrams is renormalized using  the concept of induced
interaction. The latter is implicitly defined as follows
\begin{eqnarray}
V_{ph}=V_{dir}+V_{ind}=V_{dir}+\mathcal{V}_{\rm RPA}(V_{ph}),
\label{eq-iia}
\end{eqnarray}
 where the first term $V_{dir}$ is the direct
residual interaction ($G$-matrix or Gogny force in the simplest
approximation \cite{back}) and the second one is the RPA bubble
series where, the vertex insertions are given by $V_{ph}$ itself
instead of $V_{dir}$. The construction of the induced interaction
approximation (IIA) is, in general, a very complex problem
\cite{babu,sjob,back,ains,schul}, but it is easily realized in
terms of Landau parameters, as shown in detail in the Appendix E.
In principle, the propagators and the RPA calculations should 
be calculated in the superfluid matter. However we may guess that
this consideration induces only a second order effect and we 
study the pairing screening only in the normal system in the present
paper.

\section{Results}

Based on the approximations discussed in the preceding section,
the screening interaction has been calculated for the $nn$ (or
$pp$) isospin-triplet $^{1}S_{0}$ channel and $np$ isospin-singlet
$^{3}S_{1}$ channel. In the former case the screening effects due
to both pure neutron matter and symmetric nuclear matter were also
estimated for the sake of comparison. In the past the study of
isospin-triplet pairing in nuclear matter has usually not been
considered, since one mainly had in mind the problem of
superfluidity in neutron stars, but for applications to nuclei the
investigation of pairing in nuclear matter is also relevant.

As in a previous study \cite{shen}, we used for the
calculations the Gogny D1 force (see Ref.~\cite{footnote}
and Appendix A for details) for the external vertices, as well
as for the Landau parameters describing the p-h residual
interaction in the internal vertices.

\subsection{Screening potential in the isospin triplet channel}

\begin{figure}[tbh]
\begin{center}
\includegraphics[width=7.5cm]{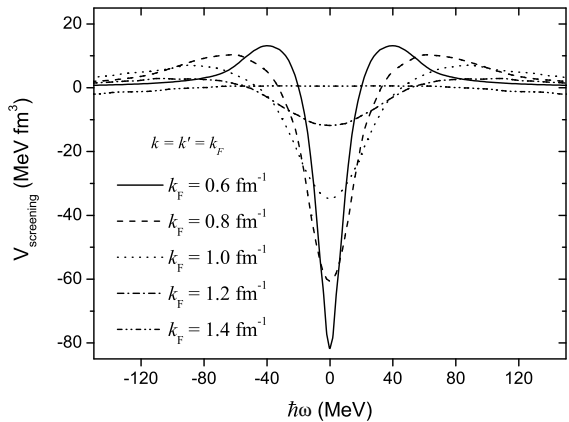}
\end{center}
\caption{Diagonal part of the sum of the neutron and proton bubble 
insertions in nuclear matter with $k_F$ ranging from 0.6 
to 1.4 fm$^{-1}$.}
\label{one-bub-np}%
\end{figure}

Let us first consider the screening of symmetric nuclear matter
and neutron matter on pairing between two nucleons in the
$^{1}S_{0}$ channel. In the one-bubble limit the neutron matter
polarization gives rise to a repulsive screening of the
$^{1}S_{0}$ $nn$ channel, as shown in Fig.~3 of Ref.\cite{shen}.
This is a well known result since long \cite{clark,ains,schulz},
and it has been interpreted as due to the dominance of the spin
density fluctuations over the density fluctuations \cite{clark0}.
It results in a sizeable quenching of the pairing gap, reinforced
by the self-energy effects \cite{shen,bald,zuo}. In nuclear
matter, the screening of $nn$ pairing in the $^1S_0$ channel can be
split into two parts: neutron bubble insertion and proton bubble
insertion, as shown in Fig.~\ref{v2-1s0-np}. The neutron bubble
produces the same effect as for $nn$ pairing in neutron matter,
but here the additional proton polarization gives a very large
attractive contribution, as shown in the right side of
Fig.~\ref{v2-1s0-np} and in Fig.~\ref{one-bub-np}. 
As a result the full medium
polarization is enhancing the isospin triplet pairing
(anti-screening), and produces, at least in the one-bubble
approximation, a huge value of the pairing gap\cite{sorb}, which
is cancelled only partially by the dispersive effects of the
self-energy treated on the same footing.

In Fig.~\ref{v2-1s0-np} we show the full 
$\omega$ dependence of the screening
part of the NN interaction for the intermediate Fermi momentum
$k_F$ = 1.0 fm$^{-1}$ and in Fig.~\ref{one-bub-np} 
we show the diagonal part of the
sum of the neutron and proton bubble insertions for $k=k'=k_F$
with $k_F$ ranging from $0.6$ to $1.4$ fm$^{-1}$. The total (i.e.
Born plus screening on shell $nn$ pairing interaction on the level
of one bubble exchange is shown in  Fig.~\ref{scree} for the $^1S_0$
channels in nuclear and neutron matter as a function of $k_F$
(dashed lines). In spite of the slight decrease of the intensity
of the induced interaction at very low nuclear matter densities
($k_F \leq$ 0.5 fm$^{-1}$), it still seems to considerably enhance
the attraction of the Born term (solid line) in the low density
limit. This qualitatively seems to be in line with the finding of
Heiselberg et al. \cite{heisel} where it was found that for a four
component Fermi system (nuclear matter) the gap should be enhanced
in the zero density limit, whereas, on the contrary, in a two
component Fermi system (neutron matter) the gap should be reduced.
This is true if the scattering lengths in all channels are
approximately equal. This is not the case in nuclear matter where
the $T=0$ scattering length is a factor three to four smaller that
the $T=1$ one and therefore a more accurate investigation of the
range $0\leq k_F \leq 0.5$ fm$^{-1}$ is still in order but
numerically very delicate.

\begin{figure}[tbh]
\begin{center}
\includegraphics[width=7cm]{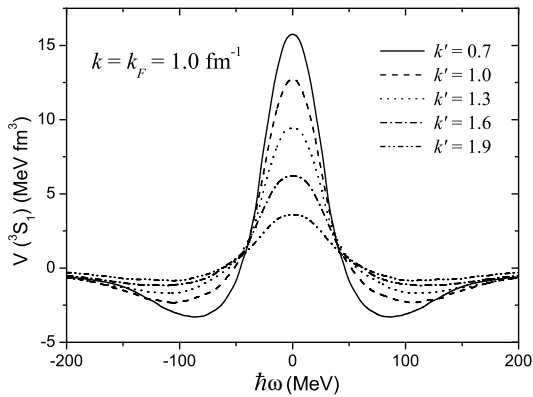}
\end{center}
\caption{Nuclear matter screening potential ${\mathcal
V}_{k_Fk'}(\omega)$ in the $^{3}S_{1}$ channel in the
one bubble approximation. The Fermi momentum $k_F$ is fixed
to 1.0 fm$^{-1}$}
\label{v2-3s1-np}%
\end{figure}

\subsection{Screening potential in the isospin singlet channel}

In nuclear matter one also has to consider the isospin-singlet
pairing, namely the pairing between unlike nucleons. The dominant
coupling interaction is due to the $^{3}S\!D_{1}$ component of the
force (in the following we neglect the $D$ component, which makes
the calculations very complicate giving however a small effect).
As mentioned above, the magnitude of the gap in this case is,
using the bare force, too large with respect to the values
observed or predicted in finite nuclei \cite{x2,bbl,bbs}. Then one
may expect that the screening is reducing the gap to a more
physical value. And in fact it turns out to be repulsive, at least
in the one-bubble approximation, as shown in Fig.~\ref{v2-3s1-np}.
Actually there are energy domains where the screening potential is
attractive, but it is repulsive around $\omega=0$, which is most
relevant for pair formation. This repulsive effect is also
confirmed for the full on shell pairing force in Fig.~\ref{scree}
especially towards lower densities (dashed line on left panel).

\subsection{Screening with induced interaction}

Regardless of the physical implications of the screening effects
on the pairing, what we learn from the above results is that the
order of magnitude of the screening or anti-screening is not small
at all. This indicates that the full bubble series (RPA) should be
summed up. On the other hand, we know that a pure RPA
approximation, namely the bubble series with the Gogny force as
residual p-h interaction, is singular in the instability region.
The only way to remove this drawback is to introduce the induced
interaction. So far the full RPA with induced interaction has been
applied to neutron matter using a $G $-matrix instead of the Gogny
force \cite{schul}. For the present calculations we adopted the
same approximation except that, as discussed in the preceding
section, the p-h residual interaction is expressed in terms of
Landau parameters. This makes the resummation of the bubble series
much easier to perform.

\begin{figure}[tbh]
\begin{center}
\includegraphics[width=85mm]{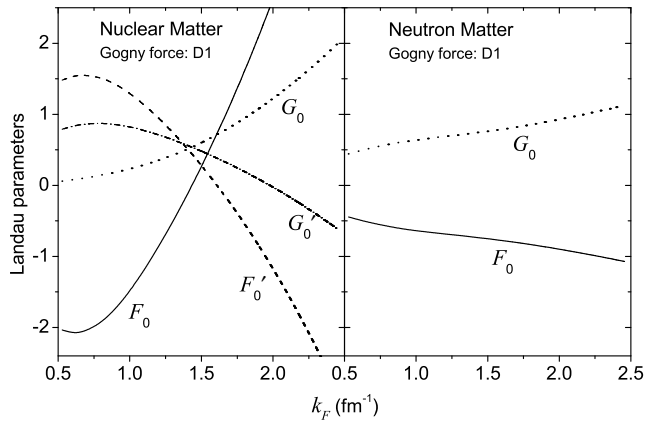}
\end{center}
\caption{Landau parameters from the Gogny force D1. Left panel
is for nuclear matter and the right one for neutron matter.}
\label{landau-0}%
\end{figure}

\begin{figure}[tbh]
\begin{center}
\includegraphics[width=85mm]{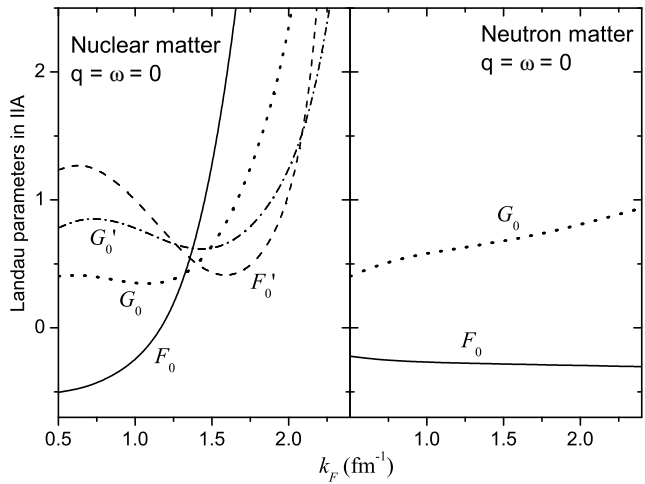}
\end{center}
\caption{The same as in Fig.~\ref{landau-0} but in the induced interaction
approximation. Only the values at the Fermi surface are depicted.}
\label{landau-iia}%
\end{figure}

\begin{figure}[tbh]
\begin{center}
\includegraphics[width=85mm]{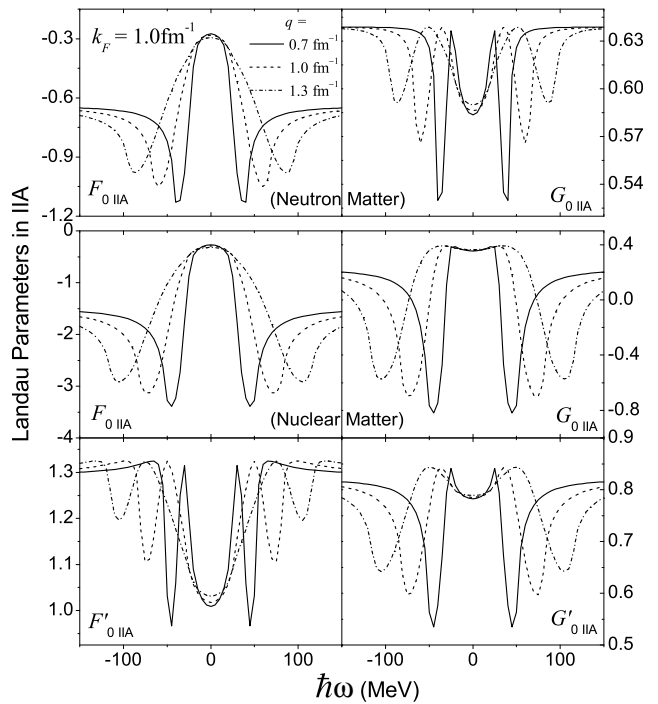}
\end{center}
\caption{Energy($\hbar \omega$) and momentum-transfer($q$) dependence
of the induced Landau parameters (see Appendix E: E2). The upper two 
panels are for neutron matter and the lower four for nuclear matter. 
The Fermi momentum $k_F$ is fixed to 1.0 fm$^{-1}$.}
\label{landau-iia-omega}
\end{figure}

\subsubsection{Landau parameters}

According to the motivations of the preceding section, the
screening interaction due to the p-h excitations is calculated
approximating the residual interaction with the $l=0$ Landau
parameters (see Ref.~\cite{bender} and Appendix B). 
The results are plotted in Fig.~\ref{landau-0} both
for nuclear matter and neutron matter. As the Brueckner
$G$-matrix, the Gogny force, where the short-range correlations
are also incorporated, does not prevent the mechanical instability
($F_{0}<-1$) to appear in the low density domain of nuclear matter
close up to the saturation point. Otherwise the values of the
Landau parameters with the Gogny force reproduce satisfactorily
the empirical values. In neutron matter $F_{0}$ slowly decreases,
since the Gogny D1 force misses the repulsive rearrangement term
in the neutron channel. The singularity due to the mechanical
instability is removed by the Babu-Brown induced interaction
\cite{babu} as shown by the enhancement of $F_{0}$ in
Fig.~\ref{landau-iia}. Since the induced interaction entails a
strong coupling among the components of the p-h interaction in
nuclear matter (see Appendix D), the rapid rise of $F_{0}$ makes
the other Landau parameters to rapidly bend up also. Despite the
simplicity of our approximation the induced interaction still
keeps dynamical effects because of its dependence on the p-h
propagator. In Fig.~\ref{landau-iia-omega} the Landau parameters
versus energy and momentum transfer in the induced interaction 
approximation, as shown in Appendix E, are plotted. The
$\hbar\omega=0$ and $q=0$ values coincide with the static values
of Fig.~\ref{landau-iia} in the induced interaction approximation,
whereas asymptotically $|\hbar\omega| \gg 0$ they tend to the zero
order limit, shown in Fig.~\ref{landau-0}. The dependence on the
momentum transfer is not significant in the range shown in the
figure, but in a wider range it could.

\subsubsection{Screening potential from the full RPA residual interaction}

\begin{figure*}[t]
\begin{center}
\includegraphics[width=120mm]{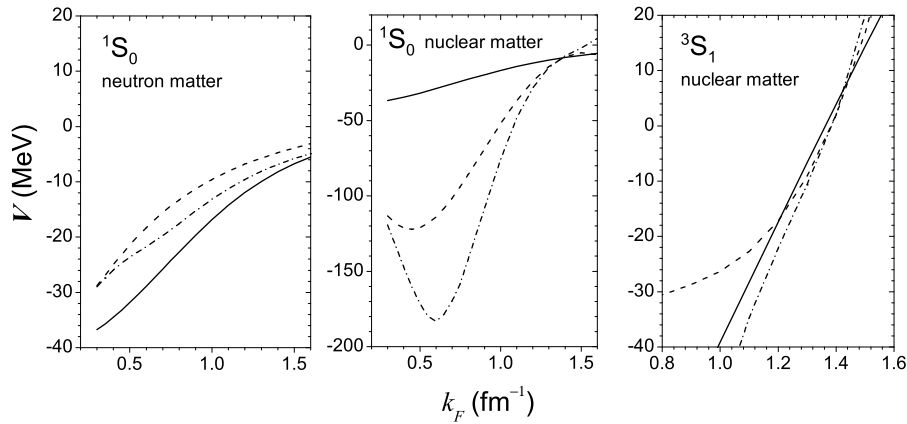}
\end{center}
\caption{Diagonal matrix elements ($k=k^{\prime}=k_{F}$) of the interaction at
$\hbar \omega=0$: Born term with Gogny force (solid lines),
Born term with one bubble insertion
(dashed lines) and full pairing potential including one and higher
bubble insertion (IIA) (dash-dotted line). The baryon density corresponds to
$k_{F} =1.0$ fm$^{-1}$. The Born term with the Gogny
force (solid line) in the $^1S_0$ channel is density independent,
whereas it is density-dependent in the $^3S_1$ channel.}
\label{scree}%
\end{figure*}

Using the Landau parameters as vertex insertions in the p-h bubble
diagrams the screening potential has been calculated in the full
RPA limit with induced interaction. The pure RPA, i.e. without
induced interaction, has not been considered, since it displays a
singular behavior in nuclear matter. Fig.~\ref{scree} summarizes
the main properties of the full pairing interaction in comparison
with the Gogny force. In the isospin triplet channel ($^{1}S_{0}$) in neutron
matter the net screening potential is repulsive because of the dominance of the
repulsive spin fluctuations over the attractive density ones.
A suppression is in accordance with practically all existing
calculations, see e.g. \cite{clark}.

For instance we also see from Fig.~\ref{one-bub-np} 
that a substantial suppression
survives at low densities. This is in agreement with the recent
studies of Gori et al.\cite{gori} and Schwenk et al. \cite{shwe}
where also an important reduction of the gap in low density
neutron matter was found.
Comparing with the low density behaviour of the one-bubble induced
interaction (see Fig.4 of Ref.[3]), we learn now that the one-bubble
approximation is insufficient. At $k_F=0.3$ fm$^{-1}$ 
its contribution to the pairing
potential is  7.93 MeV, while the bare interaction is still -36.7 MeV.
A sizeable effect is expected in the limit $N(0)V \ll 1$, which for the
nuclear force in the $T=1$ channel 
(scattering length -18.8 fm) is fulfilled
only at densities much less than $10^{-5}$ fm$^{-3}$, 
i.e., much beyond the
range of interest of the present work.

In nuclear matter the dominance of the attractive force exerted by
the proton environment on $nn$ pairs (or neutron on $pp$ pairs)
which we already found at the one bubble level, becomes more
pronounced in the full screening potential. This dominance becomes
dramatic in the case of the $^{1}S_{0}$ channel. The potential
plays a role of strong anti-screening indeed. But its effect on
the pairing correlations should be counterbalanced by the
self-energy effect which in nuclear matter are much more
pronounced than in neutron matter. The amount of compensation can
be seen in Table 1. In the case of spin triplet pairing there is a
sizeable compensation between self-energy and vertex corrections
at any density. For the spin singlet pairing in neutron matter, as
expected, both quench pair correlations, except around $k_{F}=1.0$
fm$^{-1}$. The self-energy is also reducing $^{1}S_{0}$ pairing in
nuclear matter, however, the $T=0$ part of the induced force is
enhancing the pairing interaction very much, so that a strong
(unphysical?) increase of the gap can be expected.

\begin{table}[ptb]
\caption{Comparison of dimensionless pairing interactions in three
approximations: bare interaction(first column), self-energy effects (second
column), self-energy plus screening effects (third column). ``np'' in the
second row and ``nn'' in the third row stand for nuclear matter and
neutron matter, respectively. The factor $Z$ is defined in Eq.(9) of 
Ref.~\cite{shen}.}
\begin{tabular}
[c]{c|cccc}\hline\hline
channel & \thinspace$k_{F}($fm$^{-1})$ & $N_{0}V_{\text{bare}}$ & $N_{0}%
^{\ast}Z^{2}V_{\text{bare}}$ & $N_{0}^{\ast}Z^{2}(V_{\text{bare}%
}+V_{\text{scr.}})$\\\hline
                  & 0.7 & $ -1.21 $ & $ -1.04 $ & $ -1.47 $\\
$^{3}S_{1}$       & 1.0 & $ -0.95 $ & $ -0.54 $ & $ -0.70 $\\
                  & 1.3 & $ -0.21 $ & $ -0.15 $ & $ -0.23 $\\\hline
                  & 0.7 & $ -0.43 $ & $ -0.38 $ & $ -2.46 $\\
$^{1}S_{0}($np$)$ & 1.0 & $ -0.41 $ & $ -0.23 $ & $ -1.05 $\\
                  & 1.3 & $ -0.31 $ & $ -0.22 $ & $ -0.33 $\\\hline
                  & 0.7 & $ -0.44 $ & $ -0.33 $ & $ -0.26 $\\
$^{1}S_{0}($nn$)$ & 1.0 & $ -0.41 $ & $ -0.33 $ & $ -0.26 $\\
                  & 1.3 & $ -0.31 $ & $ -0.26 $ & $ -0.21 $\\\hline\hline
\end{tabular}
\end{table}

\section{Discussion and conclusion}

In this study we included medium polarization effects in addition to the Born
term (Fig.~1a) in infinite matter in three different channels: i) $^{1}S_{0}$ pairing
in pure neutron matter, ii) $^{1}S_{0}$ pairing in symmetric nuclear matter,
iii) $^{3}S\!D_{1}$ pairing in symmetric nuclear matter. As already outlined in
the introduction, one expects from the polarization contributions (self-energy
+ vertex) quenching of pairing in i), slight additional attraction in case
ii), and quite strong quenching in iii). From Fig.~\ref{scree} we can see in
how much these expectation are realized within our calculations. We
investigated two approximations. The first just consists in considering the
one bubble exchange between two nucleons (Fig.~1b). The second also takes into account
interactions between the particle and the hole (RPA). These latter interactions, are
approximated by Landau parameters, renormalised by the Babu-Brown procedure (Fig.~1c).

The results in neutron matter for $^{1}S_{0}$ pairing confirm the
ones of other authors, i.e., the bare interaction is screened by
20 to 30\%. This screening is stronger in the one bubble
approximation than in RPA. As expected, medium polarization gives
extra attraction in the $^{1}S_{0}$ channel in symmetric nuclear
matter (case ii)). However the extra attraction turns out to be
enormous. With respect to the one bubble approximation the
resummation of the bubbles in RPA, still enhances the effect. With
such a strong total pairing force the gaps in finite nuclei would
clearly be out of scale, when applying an LDA scenario. One is
wondering what is happening. Maybe LDA is quantitatively
completely wrong when going from nuclear matter to finite nuclei.
In the past we have shown\cite{schuck,jaenicke} that for RPA, LDA
is quite good for finite momentum transfers. Here the momentum
transfers are concentrated around $q=0$ where LDA is not quite
valid. This may be the reason. A second uncertainty comes from the
theoretical approach. We see from Fig.~\ref{scree} that going from
one bubble to RPA, the effect is strong and therefore the many
body approach may not be converged. Though we very carefully
checked our formulas and the numerical procedure, an independent
confirmation of this surprisingly large effect would be indicated.
In any case it hints to the fact that also in finite nuclei
anti-screening should be taken seriously as this has indeed been
the case recently \cite{bort}.

In the $^{3}S_{1}$ channel of nuclear matter we expect quenching of
the pairing force due to medium effects. This is indeed what is happening in
the one bubble approximation. However, as in all other channels, additional
bubble summation yields extra attraction, so that with RPA also in the
$^{3}S_{1}$ channel one obtains slight extra attraction.

So let us summarize the situation. At least in the one bubble approximation
the trends in all three cases are as expected: repulsion for cases i) and
iii), attraction for case ii). In all these cases adding RPA bubble
resummation with respect to the one bubble approximation leads to considerably
more attraction. In case iii) this reverses the expected trend
from additional repulsion to additional
(slight) attraction (with respect to the Born term). Whether these additional
RPA correlations go into the right direction is not clear. Their strong effect
means that the many body scheme is not converged. An approach based on a more
controlled scheme like a variational theory would be in order. Also the
question in how much the present study can be linked to finite nuclei needs
further studies. We, however, believe that at least the trends should be the
same in nuclear matter and finite nuclei.

In the end let us again comment on our use of the Gogny force in
the Born term. As already mentioned this is not completely
consistent since for densities larger than $\approx \rho_0/4$ the
Gogny force yields a considerably larger gap than the bare NN
force (which should be used in full rigor). However, in this work
we are only interested in relative trends of the Born term versus
the induced force. Since the latter turns out to be a sizeable
fraction of the Born term, a small change of the lowest order term
willò not invalidate our conclusions of this work. On the other
hand we feel that the situation is not sufficiently under control
to warrant the huge amount of numerical work to calculate the gap
in the different channels. While in neutron matter this may yield
reasonable results which we will give in a future publication, we
can already say that in the other channels and for instance for
$T=1$ in nuclear matter the gap values will be extremely large
invalidating the whole weak coupling approach of BCS.

\begin{acknowledgments}
One of the authors (C.W. Shen) acknowledges the hospitable invitation by
INFN-LNS and Catania University. This work is partly supported by the
Natural Science Foundation of China (Grant No. 10305019, 10235020) and the
Chinese Academy of Science Knowledge Innovation Project
(Grant No. KJCX1-N11).
\end{acknowledgments}


\appendix

\section{Gogny force}

The appendix is devoted to develop the formalism describing the interaction
and the approximations adopted in the calculations. The Gogny force,
neglecting the spin-orbit term, is given by Ref.~\cite{gogny}
\begin{eqnarray}
\left\langle k_{1}k_{2}|V|k_{3}k_{4}\right\rangle &=&
\sum_{i=1,2}\mu_i e^{-r_{i}^{2}(\bm{k}_{1}-\bm{k}_{3}
)^{2}/4}(W_{i}+B_{i}P^{\sigma}\nonumber\\
&&  -H_{i}P^{\tau}-M_{i}P^{\sigma}P^{\tau})\nonumber \\
&& +\gamma (1+P^\sigma),
\end{eqnarray}
where $k\equiv(\bm{k},\sigma,\tau)$ denotes the single-particle state and
$\mu$, $\gamma$ is defined as
\begin{equation}
\mu_{i}=(\sqrt{\pi}r_{i})^{3},\ \ \ \ \gamma=t_{3}\rho^{1/3},
\label{app-a1}
\end{equation}
which will be used in Appendix B, C, D. The parameters
of Gogny D1 force are reported in Table I.
\begin{table}[tbh]
\caption{The parameters D1 for Gogny force. \newline($t_{3}=1350$ MeV fm$^{4}%
$)}%
\begin{ruledtabular}
\begin{tabular}{c|ccccc}
& $r_{i}$(fm) & $W_{i}$ & $B_{i}$ & $H_{i}$ & $M_{i}$(MeV) \\
\hline
$1$ & $0.7$ & $-402.4$ & $-100$ & $-496.2$ & $-23.56$ \\
$2$ & $1.2$ & $-21.3$ & $-11.77$ & $37.27$ & $-68.81$ \\
\end{tabular}%
\end{ruledtabular}
\end{table}

\section{Landau parameters from Gogny force}

The p-h residual interaction can be described in terms of the
Landau parameters to be extracted from the preceding Gogny force.
In order to do that we introduce the parameters
\begin{equation}
n_{\bm{k}}^{ST}=\sum_{\sigma,\tau}\,c_{\sigma,\tau}^{ST}\,n_{\bm{k}%
\sigma,\tau}%
\end{equation}
where
\[
c_{\sigma\tau}^{ST}=\left\{
\begin{array}
[c]{ll}%
1 & \text{baryon density (}S\text{=0,}T\text{=0)},\\
(-1)^{\sigma-\frac{1}{2}} & \text{spin density (}S\text{=1,}T\text{=0)},\\
(-1)^{\tau-\frac{1}{2}} & \text{isospin density (}S\text{=0,}T\text{=1)},\\
(-1)^{\sigma+\tau-1} & \text{spin-isospin density (}S\text{=1,}T\text{=1)}.%
\end{array}
\right.
\]
Spin and isospin asymmetric nuclear matter can be is expressed in terms of the
previous parameters as follows
\begin{equation}
\mathcal{H}=\mathcal{K}+\frac{1}{2}\sum_{kk^{\prime}}\left\langle kk^{\prime
}|\bar{V}|kk^{\prime}\right\rangle n_{k}^{\prime}n_{k^{\prime}},
\end{equation}
where $\mathcal{K}$ is the kinetic part and $\bar{V}$ denotes anti-symmetrized
interaction. Then the Landau parameters are obtained as functional
derivatives
\begin{equation}
F_{\bm{k}\bm{k}^{\prime}}^{ST}=N(0)f_{\bm{k}\bm{k}^{\prime}}%
^{ST}=N(0)\frac{\delta^{2}\mathcal{H}}{\delta n_{\bm{k}}^{ST}\delta
n_{\bm{k}^{\prime}}^{ST}}.%
\end{equation}
The normalization factor N(0) is the level density and it is introduced to
make the Landau parameters to be dimensionless. After expanding into Legendre
polynomials, we select for our purpose only the zero-order Landau parameters.
For nuclear matter we get
\begin{eqnarray}
F_{0}  &  =&\sum_{i}[(4W_{i}+2B_{i}-2H_{i}-M_{i})u_{i}-(W_{i}+2B_{i}\nonumber\\
         &&  -2H_{i}-4M_{i})v_{i}]+\tfrac{7}{6}\gamma,\nonumber\\
F_{0}^{\prime}  &  =&\sum_{i}[(-W_{i}-2B_{i})u_{i}-(2H_{i}+M_{i})v_{i}%
]-\tfrac{3}{4}\gamma,\nonumber\\
G_{0}  &  =&\sum_{i}[(2B_{i}-M_{i})u_{i}-(W_{i}-2H_{i})v_{i}]+\tfrac{1}{4}%
\gamma,\nonumber\\
G_{0}^{\prime}  &  =&\sum_{i}[-M_{i}u_{i}-W_{i}v_{i}]-\tfrac{1}{4}
\gamma,
\end{eqnarray}
where $u_{i}=\mu_{i}N(0)/4$ and $v_{i}=u_{i}e^{-z_{i}}(\sinh z_{i})/z_{i}$ and
$z_{i}=(r_{i}k_{F})^{2}/2$ and $|\bm{k}|=|\bm{k}^{\prime}|=k_{F}$, $k_{F}$
being the Fermi momentum. For nuclear matter $N(0)=2m^{\ast}k_{F}/(\pi
\hbar)^{2}$. $\gamma$ and $\mu_i$ are defined in Eq.~(A2).

In the case of pure neutron matter the zero range term of the Gogny force does
not contribute. Only two (zero order) Landau parameters survive, i.e.
\begin{eqnarray}
F_{0}  &  =&\sum_{i}[(2W_{i}+B_{i}-2H_{i}-M_{i})u_{i}\nonumber\\
        &&  -(W_{i}+2B_{i}-H_{i}-2M_{i})v_{i}],\nonumber\\
G_{0}  &  =&\sum_{i}[(B_{i}-M_{i})u_{i}-(W_{i}-H_{i})v_{i}],
\end{eqnarray}
where now $u_{i}$=$\mu_{i}N(0)/2$ and $N(0)$=$m^{\ast}k_{F}/(\pi\hbar)^{2}$ is
the level density of neutron matter, $k_{F}$ being the neutron-matter Fermi momentum.

\section{One-bubble screening interaction}

The medium polarization effects on the NN interaction in the RPA limit are
given by the bubble series. The one-bubble term (diagram) has the following expression,
\begin{equation}
V^{(2)}=-\sum_{k_{1}k_{2}}\left\langle kk_{1}|\bar{V}|k^{\prime}%
k_{2}\right\rangle \left\langle k_{2}\bar{k}|\bar{V}|k_{1}\bar{k}^{\prime
}\right\rangle L_{k_{1}k_{2}}(\omega), \label{app-5}%
\end{equation}
which is suitable for the pairing interaction where one particle in the state
$k\equiv(\bm{k},\sigma,\tau)$ couples to one particle in the state $\bar
{k}\equiv(-\bm{k},\sigma^{\prime},\tau^{\prime})$. The function $L $ is the
polarization part~\cite{fet}
\begin{equation}
L_{k_{1}k_{2}}(\omega)=\int\frac{d\omega^{\prime}}{2\pi i}\tilde{G}_{k_{1}%
}(\omega^{\prime})G_{k_{2}}(\omega-\omega^{\prime}).
\end{equation}
After $\omega$-integration within the single-pole approximation for the
Green's function, one obtains
\[
L_{k_{1}k_{2}}(\omega)=\frac{n_{k_{2}}(1-n_{k_{1}})Z_{k_{2}}}{\omega_{k_{1}%
}+\omega_{k_{2}}+\omega-i\eta}+\frac{n_{k_{1}}(1-n_{k_{2}})Z_{k_{1}}}%
{\omega_{k_{1}}+\omega_{k_{2}}-\omega-i\eta},
\]
where the $Z$-factors embody the self-energy effects. In the coupled case,
Eq.(\ref{app-5}) becomes,
\begin{equation}
V_{\text{ST}}^{(2)}(k,k^{\prime},\omega)=-\frac{1}{4\pi}\int\hspace{-2mm}%
\sum\limits_{\bm{k}_{1}\bm{k}_{2},ij}V_{ij}^{(2)}(\text{ST})L_{k_{1}k_{2}%
}(\omega)d\Omega_{\bm{k}}d\Omega_{\bm{k}^{\prime}},
\end{equation}
where ST stands for $^{1}S_{0}$ or $^{3}S_{1}$ channel, and $i,j$ takes 1 and
2. In the neutron matter medium,
\begin{eqnarray}
&&  V_{ij}^{(2)}(^{1}S_{0})\nonumber\\
=\hspace{-3mm}  &&  2(\alpha_{i}-\beta_{i})(B_{i}-H_{i}-M_{i}+W_{i})[\beta
_{j}(H_{j}-W_{j})\nonumber\\
&&  +\eta_{j}(B_{j}-M_{j})]+[\alpha_{i}(H_{i}-W_{i})+\beta_{i}(B_{i}-M_{i})]\nonumber\\
&&  \times\lbrack\beta_{j}(M_{j}-B_{j})+\eta_{j}(W_{j}-H_{j})],
\end{eqnarray}
where $\alpha_{i}$=$\mu_{i}$exp$[-\frac{1}{4}r_{i}^{2}(\bm{k}-\bm{k}%
_{2})^{2}]$, $\beta_{i}$=$\mu_{i}$exp$[-\frac{1}{4}r_{i}^{2}(\bm{k}-\bm
{k}^{\prime})^{2}]$, $\eta_{i}$=$\mu_{i}$exp$[-\frac{1}{4}r_{i}^{2}(\bm
{k}-\bm{k}_{1})^{2}]$, and in nuclear matter medium,
\begin{eqnarray}
&&  V_{ij}^{(2)}(^{1}S_{0})\nonumber\\
=\hspace{-3mm}  &&  -(\alpha_{i}H_{i}+\beta_{i}B_{i}+\tfrac{1}{2}\gamma)(\beta_{j}%
B_{j}+\eta_{j}H_{j}+\tfrac{1}{2}\gamma)\nonumber\\
&&  +2(\alpha_{i}-\beta_{i})(B_{i}-H_{i}-M_{i}+W_{i})[\beta_{j}(H_{j}-W_{j})\nonumber\\
&&  +\eta_{j}(B_{j}-M_{j})]+2[\alpha_{i}(H_{i}+M_{i})+\beta_{i}(B_{i}+W_{i})\nonumber\\
&&  +\gamma][\beta_{j}W_{j}+\eta_{j}M_{j}+\tfrac{1}{2}\gamma]-[\alpha_{i}(W_{i}-H_{i})\nonumber\\
&&  +\beta_{i}(M_{i}-B_{i})][\beta_{j}(M_{j}-B_{j})+\eta_{j}(W_{j}-H_{j})].
\end{eqnarray}
In $^{3}S_{1}$ channel,%
\begin{eqnarray}
&&  V_{ij}(^{3}S_{1})\nonumber\\
=\hspace{-3mm}  &&  -(\alpha_{i}B_{i}+\beta_{i}H_{i}+\tfrac{1}{2}\gamma)(\beta_{j}%
H_{j}+\eta_{j}B_{j}+\tfrac{1}{2}\gamma)\nonumber\\
&&  +2[\alpha_{i}(M_{i}-B_{i})+\beta_{i}(W_{i}-H_{i})](\beta_{j}W_{j}+\eta
_{j}M_{j}\nonumber\\
&&  +\tfrac{1}{2}\gamma)+2(\alpha_{i}-\beta_{i})(M_{i}-W_{i}+H_{i}-B_{i})[\beta_{j}%
(B_{j}\nonumber\\
&&  +W_{j})+\eta_{j}(H_{j}+M_{j})+\gamma]-[\alpha_{i}(B_{i}+W_{i})+\beta
_{i}(H_{i}\nonumber\\
&&  +M_{i})+\gamma][\beta_{j}(H_{j}+M_{j})+\eta_{j}(B_{j}+W_{j})+\gamma
],
\end{eqnarray}
where $\gamma$ is density dependent, as defined in Eq.(\ref{app-a1}).

\section{RPA screening interaction}

The summation of the full RPA bubble series can be carried out
only in few cases \cite{nav}, as explained in Sec.~II. A reasonable
approximation which preserves the correct evaluation of vertex
insertions between p-p external lines and the p-h internal lines is
depicted in Fig.~1(c). It is based on a two-bubble approximation where
the internal vertex is replaced by the full RPA p-h interaction
\begin{eqnarray}
V^{(3)} &=&\sum_{k_{1}k_{2}k_{3}k_{4}}
\left\langle kk_{1}\right|\bar{V}\left|k^{\prime
}k_{2}\right\rangle L_{\bm{k}_{1}\bm{k}_{2}}(\omega)\nonumber\\
&&  \times\langle k_{2}k_{3}|\tilde{V}_{\rm RPA}|k_{1}k_{4}\rangle
L_{\bm{k}_{3}\bm{k}_{4}}(\omega)\left\langle k_{4}\bar{k}|\bar{V}|k_{3}%
\bar{k}^{\prime}\right\rangle. \nonumber \\
&&
\end{eqnarray}
The momentum transfer in the pairing interaction is $\bm{q}\equiv\bm{k}%
-\bm{k}^{\prime}=\bm{k}_{2}-\bm{k}_{1}=\bm{k}_{4}-\bm{k}_{3}$.
Using the Landau parameters, $\tilde{V}_{\rm RPA}$ is only depending
on $\bm{q}$. The latter is calculated using the Landau parameters
for which the RPA summation can be easily performed. In turn, this
is a reasonable approximation, since the relevant p-h excitations
are those nearby the Fermi surface where
the Landau parameters are calculated. Therefore, in the Landau approximation,%
\begin{equation}
\tilde{V}_{\rm RPA}=\tilde{f}+\tilde{g}\boldsymbol{\sigma}_{1}\cdot
\boldsymbol{\sigma}_{2}+\tilde{f}^{\prime}\boldsymbol{\tau}_{1}\cdot
\boldsymbol{\tau}_{2}+\tilde{g}^{\prime}(\boldsymbol{\sigma}_{1}%
\cdot\boldsymbol{\sigma}_{2})(\boldsymbol{\tau}_{1}\cdot\boldsymbol{\tau}%
_{2}),
\end{equation}
where $\tilde{f}$, $\tilde{g}$, $\tilde{f}^{\prime}$, $\tilde{g}^{\prime}$,
including so called induced interaction, will be given in Appendix E.\ In the
coupled case, the interaction with full RPA (the bubble series summation with
two and higher bubbles) is
\begin{equation}
V_{ST}^{(3)}=\frac{1}{4\pi}\int d\Omega_{k}d\Omega_{k^{\prime}}\sum
\limits_{\bm{k}_{2}\bm{k}_{4},ij}V_{ij}^{(3)}(ST)
\end{equation}
where $ST$ could be $^{1}S_{0}$ or $^{3}S_{1}$. For $^{1}S_{0}$ channel,
$V_{ij}$ takes the form%
\begin{eqnarray}
V_{ij}^{(3)}  &  =&(P_{i}\alpha_{i}+Q_{i}\beta_{i})(P_{j}\alpha_{j}+Q_{j}%
\eta_{j})\tilde{f}\nonumber\\
&&  -3(X_{i}\alpha_{i}+Y_{i}\beta_{i})(X_{j}\alpha_{j}+Y_{j}\eta_{j})\tilde{g}
\end{eqnarray}
in neutron matter, where $\alpha_{i}=\mu_{i}{\rm exp}[-r_{i}^{2}(\bm{k}-\bm
{k}^{\prime})^{2}/4]$, $\beta_{i}=\mu_{i}{\rm exp}[-r_{i}^{2}(\bm{k}-\bm{k}%
_{2})^{2}/4]$, $\eta_{i}=\mu_{i}{\rm exp}[-r_{i}^{2}(\bm{k}+\bm{k}_{4})^{2}/4]$,
$P_{i}=B_{i}-2H_{i}-M_{i}+2W_{i}$, $Q_{i}=-2B_{i}+H_{i}+2M_{i}-W_{i}$,
$X_{i}=B_{i}-M_{i}$, $Y_{i}=H_{i}-W_{i}$, and
\begin{eqnarray}
V_{ij}\hspace{-2mm}^{(3)}\hspace{-2mm}  & =&\hspace{-2mm}(A_{i}\alpha
_{i}-E_{i}\beta_{i}+\tfrac{3}{2}\gamma)(A_{j}\alpha_{j}-E_{j}\eta_{j}+\tfrac
{3}{2}\gamma)(\tilde{f}-\tilde{f}^{\prime})\nonumber\\
&&  \hspace{-4mm}-3(C_{i}\alpha_{i}+D_{i}\beta_{i}+\tfrac{1}{2}\gamma)(C_{j}\alpha
_{j}+D_{j}\eta_{j}+\tfrac{1}{2}\gamma)(\tilde{g}-\tilde{g}^{\prime})\nonumber \\
&&
\end{eqnarray}
in nuclear matter where $A_{i}=2B_{i}-2H_{i}-M_{i}+4W_{i}$, $E_{i}%
=2B_{i}-2H_{i}-4M_{i}+W_{i}$, $C_{i}=2B_{i}-M_{i}$, $D_{i}=2H_{i}-W_{i}$. For
$^{3}S_{1}$ channel, we have%
\begin{eqnarray}
V_{ij}\hspace{-2mm}^{(3)}\hspace{-2mm}  & =&\hspace{-2mm}(A_{i}\alpha
_{i}-E_{i}\beta_{i}+\tfrac{3}{2}\gamma)(A_{j}\alpha_{j}-E_{j}\eta_{j}+\tfrac
{3}{2}\gamma)(\tilde{f}-\tilde{f}^{\prime})\nonumber\\
&&  \hspace{-4mm}+(C_{i}\alpha_{i}+D_{i}\beta_{i}+\tfrac{1}{2}\gamma)(C_{j}\alpha
_{j}+D_{j}\eta_{j}+\tfrac{1}{2}\gamma)(\tilde{g}-\tilde{g}^{\prime}).\nonumber\\
&&
\end{eqnarray}

\section{Induced interaction}

The induced interaction is given by Eq.(\ref{eq-iia}), i.e.
\begin{equation}
\tilde{V}_{\rm RPA}=V_{dir}+\mathcal{V}_{\text{ind}}.
\end{equation}
where the induced interaction $\mathcal{V}_{ind}$ is the bubble series with
the full interaction itself inserted in the interaction vertices. Its explicit
expression, in nuclear matter case, is,%
\begin{eqnarray}
\tilde{f}  &  =&f_{d}+f_{\text{ind}}\text{, \ \ \ }\tilde{f}^{\prime}%
=f_{d}^{\prime}+f_{\text{ind}}^{\prime},\nonumber\\
\tilde{g}  &  =&g_{d}+g_{\text{ind}}\text{, \ \ }\tilde{g}^{\prime}%
=g_{d}^{\prime}+g_{\text{ind}}^{\prime}, \label{app-1}%
\end{eqnarray}
where%
\begin{eqnarray}
4f_{\text{ind}}  &  =&A_{f}+3A_{g}+3A_{f^{\prime}}+9A_{g^{\prime}},\nonumber\\
4g_{\text{ind}}  &  =&A_{f}-A_{g}+3A_{f^{\prime}}-3A_{g^{\prime}},\nonumber\\
4f_{\text{ind}}^{\prime}  &  =&A_{f}+3A_{g}-A_{f^{\prime}}-3A_{g^{\prime}%
},\nonumber\\
4g_{\text{ind}}^{\prime}  &  =&A_{f}-A_{g}-A_{f^{\prime}}+A_{g^{\prime}},
\label{app-2}%
\end{eqnarray}
and%
\begin{eqnarray}
A_{f}  &  =&\frac{\Lambda}{1+N(0)\tilde{f}\Lambda}\tilde{f}^{2},\ \
A_{f^{\prime}}=\frac{\Lambda}{1+N(0)\tilde{f}^{\prime}\Lambda}\tilde{f}%
^{\prime2},\nonumber\\
A_{g}  &  =&\frac{\Lambda(q)}{1+N(0)\tilde{g}\Lambda}\tilde{g}^{2},\ \
A_{g^{\prime}}=\frac{\Lambda(q)}{1+N(0)\tilde{g}^{\prime}\Lambda}\tilde
{g}^{\prime2}. \label{app-3}%
\end{eqnarray}
While in neutron matter case, only $\tilde{f}$ and $\tilde{g}$ is remained in
Eq.(\ref{app-1}) and the self-contained equation is%
\begin{eqnarray}
2f_{\text{ind}}  &  =&A_{f}+3A_{g},\nonumber\\
2g_{\text{ind}}  &  =&A_{f}-A_{g}, \label{app-4}%
\end{eqnarray}
where $A_{f}$ and $A_{g}$ has the same form as in Eq.(\ref{app-2}). Usually
Eq.(\ref{app-1}) and (\ref{app-4}) is solved iteratively, in which the first
guess is the direct term $f_{d}$. The simplest approximation to calculate the
induced interaction is given expressing $f$ in terms the $L=0$ Landau
parameters where $\Lambda=\Lambda(q,0)$ is the Lindhard function in the static
limit $\omega=0$.

\end{document}